\documentclass[twocolumn]{aastex62}

\newcommand{\kms}{\,km\,s$^{-1}$}

\newcommand{\Msun}{$M_{\odot}$}

\newcommand{\Msunyr}{$M_{\odot}$yr$^{-1}$}

\newcommand{\rev}{ }

\newcommand{\sect}{Sect.\,}

\newcommand{\OIa}{[O\,{\scriptsize I}]\,$\lambda$6300}
\newcommand{\OIb}{[O\,{\scriptsize I}]\,$\lambda$5577}

\def\arcsec{\hbox{$^{\hbox{\rlap{\hbox{\lower4pt\hbox{$\,\prime\prime$}}}\hbox{$\frown$}}}$}}
          
\graphicspath{{./}{figures/}}

\shorttitle{}
\shortauthors{}

\begin{document}

\title{Double-peaked OI profile: a likely signature of the gaseous ring around KH~15D}

\author{Min Fang}
\affiliation{Department of Astronomy, University of Arizona, 933 North Cherry Avenue, Tucson, AZ 85721, USA}
\affiliation{Earths in Other Solar Systems Team, NASA Nexus for Exoplanet System Science}
\author{Ilaria Pascucci}
\affiliation{Department of Planetary Sciences, University of Arizona, 1629 East University Boulevard, Tucson, AZ 85721, USA}
\affiliation{Earths in Other Solar Systems Team, NASA Nexus for Exoplanet System Science}
\author{Jinyoung Serena Kim}
\affiliation{Department of Astronomy, University of Arizona, 933 North Cherry Avenue, Tucson, AZ 85721, USA}
\affiliation{Earths in Other Solar Systems Team, NASA Nexus for Exoplanet System Science}
\author{Suzan Edwards}
\affiliation{Five College Astronomy Department, Smith College, Northampton, MA 01063, USA}



\begin{abstract}
  KH~15D is a well-known spectroscopic binary because of its unique and dramatic photometric variability.
 The variability is explained by a circumbinary  dust ring but the ring itself was never directly detected. We present a new interpretation of the double-peaked \OIa\ profiles as originating from the  hot disk surface of KH~15D.  By modeling these profiles, we measure emitting radii between $\sim$0.5--5\,au,  basically a gaseous ring very similar in radial extent to the dust ring inferred from modeling the system's photometric variability. We also discuss the possibility that external photoevaporation driven by UV photons from the nearby massive star HD~47887 has truncated the outer edge of the disk to the observed value.

\end{abstract}

\keywords{open clusters and associations: individual (NGC 2264) --- protoplanetary disks --- stars: individual (KH 15D) --- stars: pre-main sequence}



\section{Introduction} \label{sec:introduction}

KH~15D is a pre-main sequence (PMS) spectroscopic binary (A+B) with near-equal-mass components (0.715\Msun\ for Star A and 0.74\Msun\ for Star B), high eccentricity ($e\approx 0.6$), a semi-major axis of $\sim$0.25~au, and a period of 48.37 days   \citep{2004AJ....128.1265J,2006ApJ...644..510W,2018AJ....155...47A}.  It is a member of the NGC~2264 cluster which is at $\sim750$\,pc, based on the members from \citealt{2018A&A...609A..10V} and \textit{Gaia} Data Release 2 parallaxes, and the cluster is about $\sim$2--6\,Myr old \citep{2009MNRAS.399..432N,2016ApJ...831..116L}.  KH~15D exhibits unique and dramatic long and short term photometric variability (e.g., \citealt{2012ApJ...757L..18C}). {\rev  From 2006 to 2009 Star B was fully occulted while after 2010 Star A became occulted; currently, the entirety of Star B is visible at each apastron passage  \citep{2018AJ....155...47A}}
The short-term variability has the same period {\rev as} the binary
 \citep{1998AJ....116..261K,2001ApJ...554L.201H,2002PASP..114.1167H,2004AJ....128.1265J}. However, the large depth (several magnitudes) and long duration (days to tens {\rev of} days) of the eclipse, in combination with the time variability,  cannot be solely attributed to the binary. Additional occultation by circumbinary matter must be invoked \citep{2001ApJ...554L.201H,2002PASP..114.1167H}.

\begin{figure}
\begin{center}
\includegraphics[width=\columnwidth]{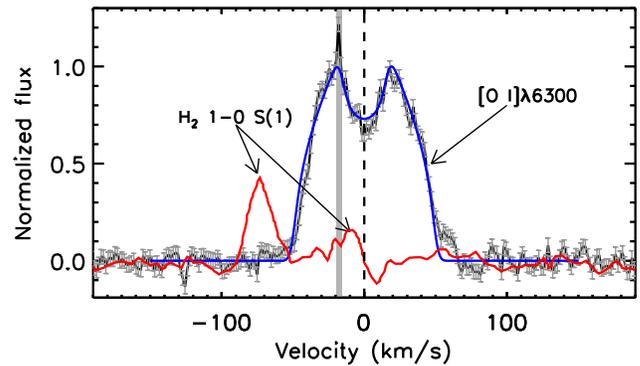}
\caption{\OIa\ line profile (grey line) in the system's reference frame obtained on 2001 December 21. The profile is compared with the  H$_{2}$\,1-0\,S(1) digitally extracted from Figure~2 of \cite{2004ApJ...601L..87D} (red line). The best fit model profile  is also shown in the panel (blue line). The grey-color filled region marks the wavelength range  contaminated by the sky line and  thus excluded in the fitting (see the discussion in Sect.~\ref{sec:data}).  
}\label{Fig:meanline_profile}
\end{center}
\end{figure}

A nearly edge-on, warped and precessing {\rev circumbinary ring of dust} ($\sim$1 to 5~au)   is thought to occult the eccentric PMS binary and explain the available light curves   \citep{2004ApJ...607..913C,2004ApJ...603L..45W}. However, the circumbinary disk eluded detection. The system shows no infrared excess emission out to 8\,$\mu m$ \citep{2016AJ....151...90A},  in line with the relatively large inner dust edge. In addition, the  dust ring has not been detected at millimeter wavelengths with SMA  or ALMA.  The millimeter observations have  yielded upper limits on the total (gas+dust) circumbinary disk mass of  1.7~$M_{\rm JUP}$ \citep{2018AJ....155...47A},  just a factor of $\sim 2$ lower than the mean protoplanetary disk mass in young (1-3\, Myr) star-forming regions (e.g., \citealt{2016ApJ...828...46A}, \citealt{2016ApJ...831..125P}). Interestingly, the SMA CO~$J=3-2$  line reveals a bipolar collimated outflow, the northern lobe of which is spatially coincident with the  H$_{2}$ jet associated with KH~15D \citep{2004ApJ...601L..91T,2018AJ....155...47A}.  Finally, the system presents broad H$\alpha$ line profiles, indicative of  active accretion \citep{2012ApJ...751..147H}, perhaps funneled from gas surrounding both stars (see the case of the spectroscopic binary DQ~Tau, Muzerolle et al. 2019). The presence of a jet and ongoing accretion point to the presence of a gaseous disk.
 
 Here, we interpret the double-peaked \OIa\ profiles from  KH~15D as originating from the surface of  this gaseous disk ($\S$2). Modeling of the line profiles reveals that the  emitting gas is radially confined ($\S$3). We discuss how external photoevaporation could have truncated  the gaseous disk to the outer edge inferred from our \OIa\ modeling  ($\S$4).

\begin{figure*}
\begin{center}
\includegraphics[width=1.\columnwidth]{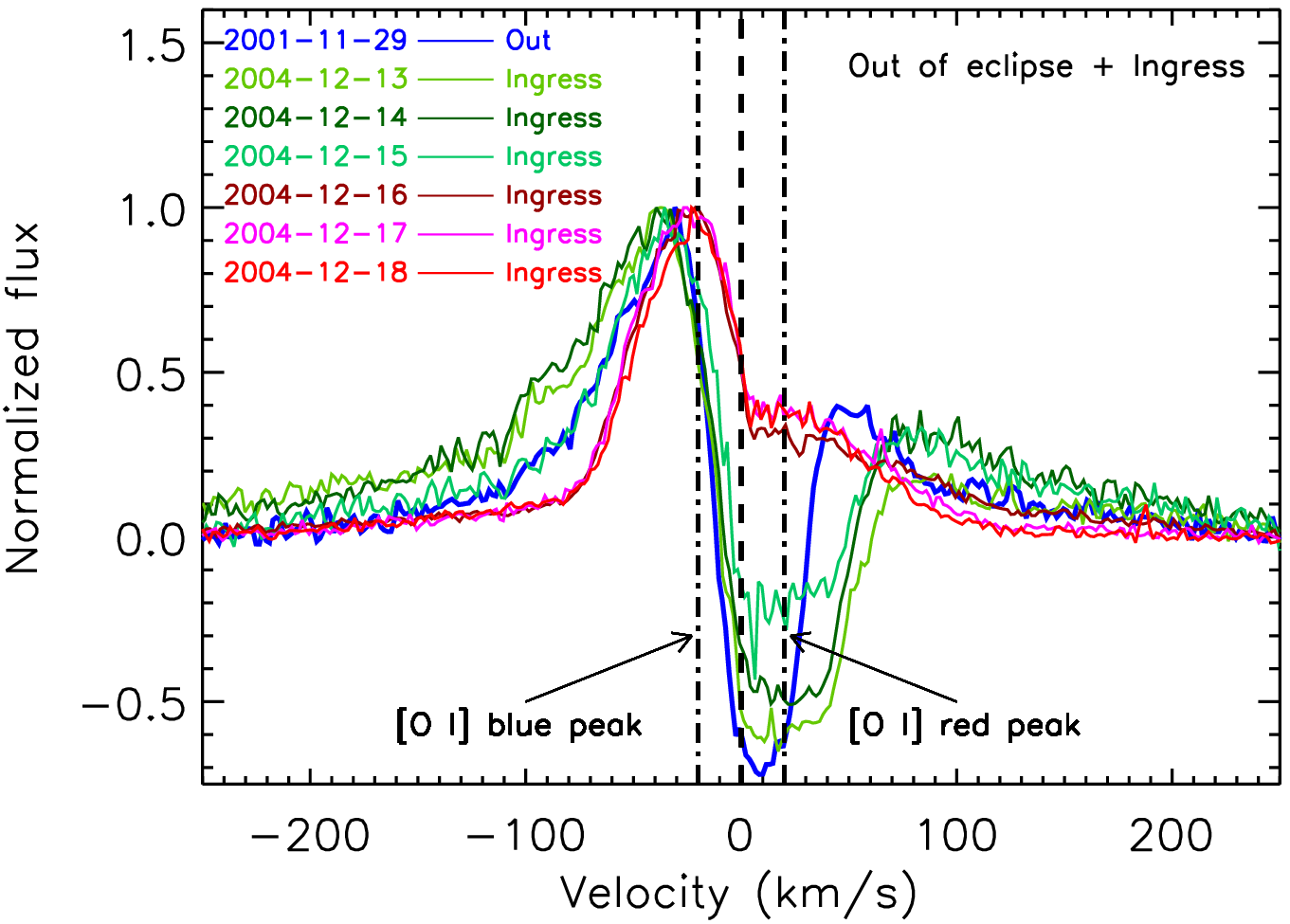}
\includegraphics[width=1.\columnwidth]{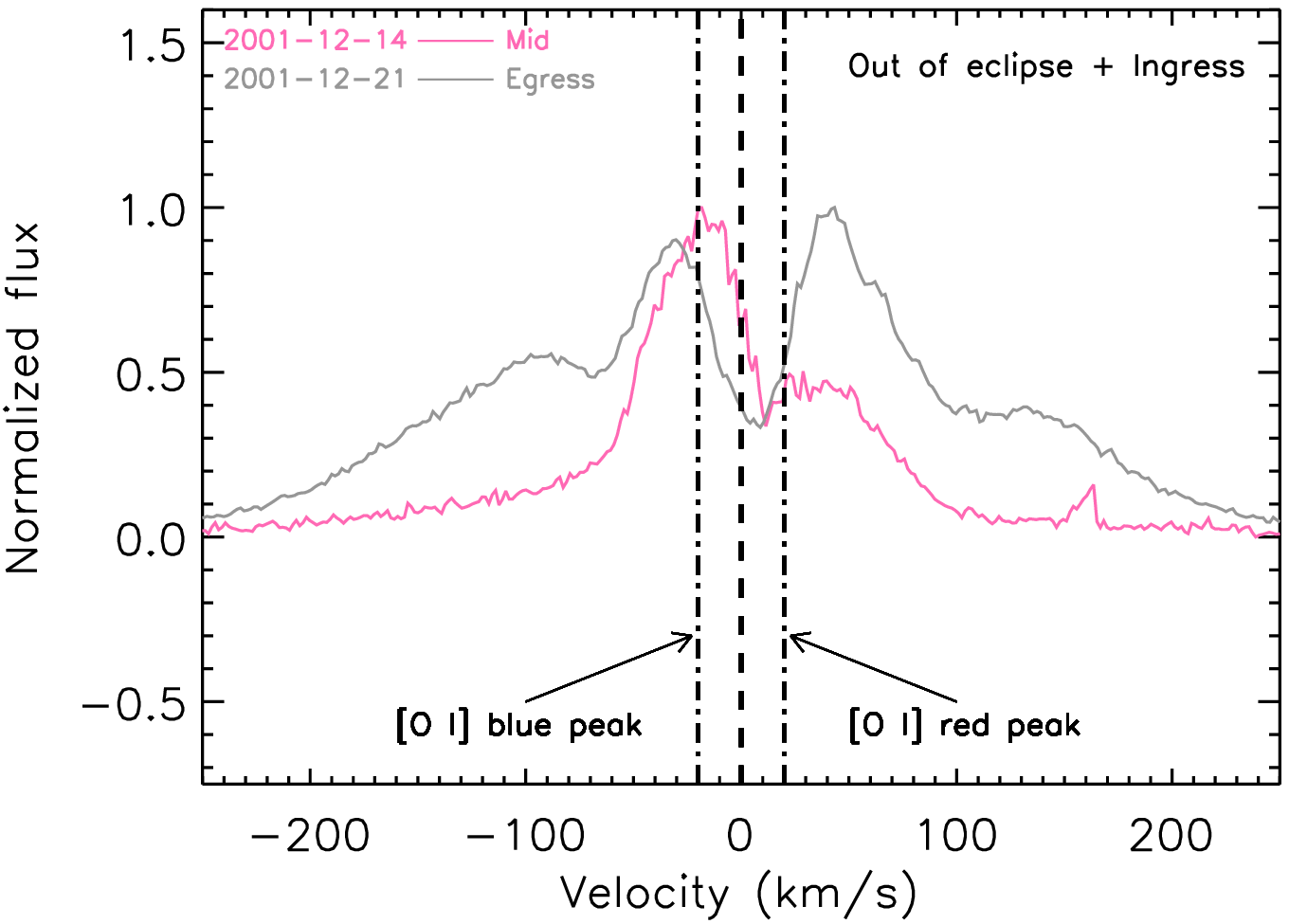}
\caption{Left: H$\alpha$ line profiles when star A is fully occulted by the disk. Right: H$\alpha$ line profiles when star A is partially occulted by the disk or out of the eclipse. Profiles are in the system's reference frame.  The dash-dotted lines mark the two peaks ($\pm$20\kms) from the \OIa\ profiles. Note that the H$\alpha$ profiles shown here and \OIa\ profiles in Fig.~\ref{Fig:line_profile}  are extracted from the same spectra, hence correspond to the same observing times. }\label{Fig:line_profile_Halpha}
\end{center}
\end{figure*}

\begin{figure*}
\begin{center}
\includegraphics[width=1.5\columnwidth]{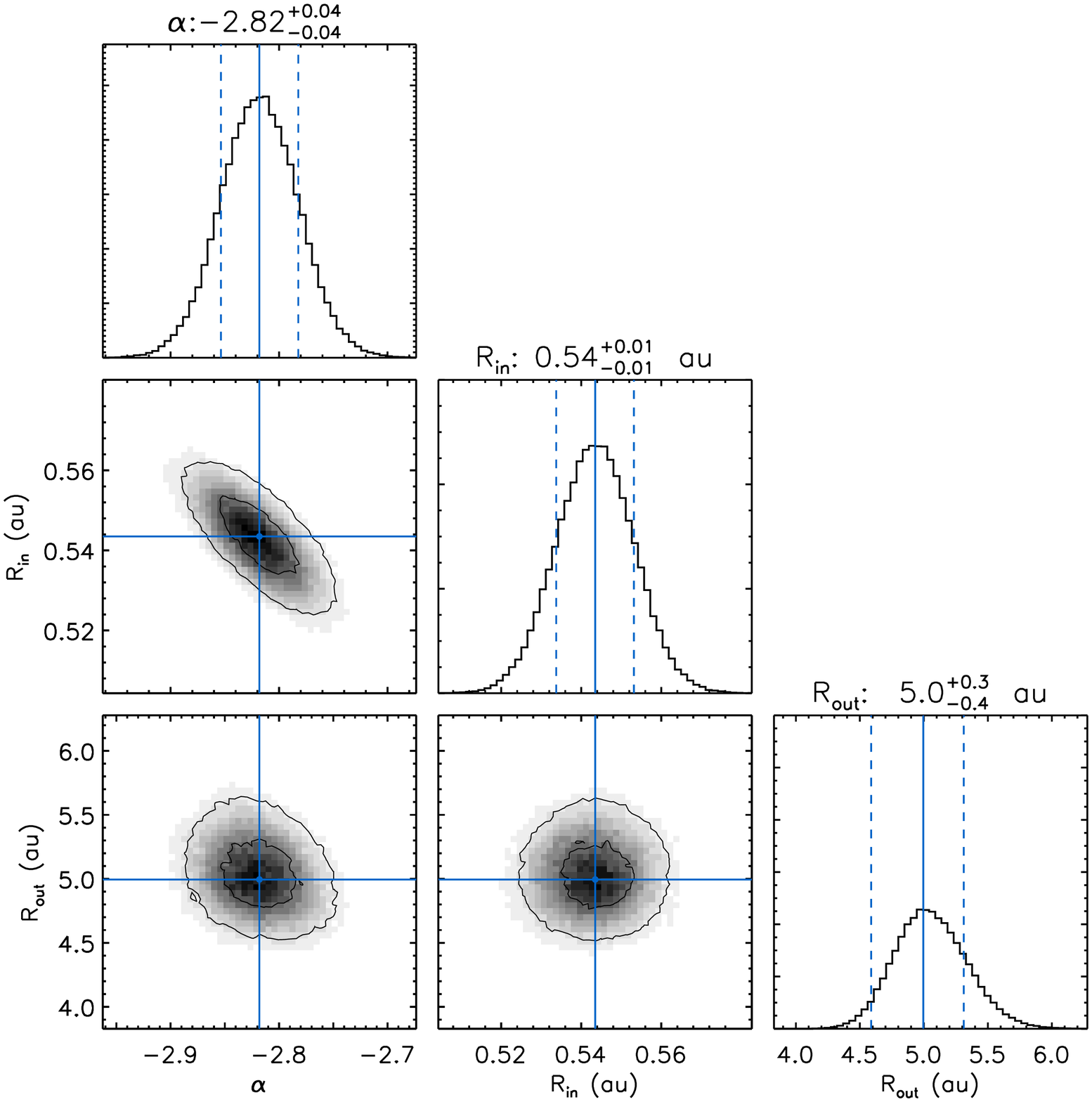}
\caption{Corner plot showing the posterior distributions from the MCMC fit of the \OIa\ line obtained on 2001 December 21 (see Fig.~\ref{Fig:meanline_profile}). The vertical dashed lines are the 16 and 85 percentiles, respectively. The solid lines indicate the medians of the posterior distributions.}\label{Fig:Corner}
\end{center}
\end{figure*}

\section{Double peaked \OIa\ profiles} \label{sec:data}
The spectroscopic data  presented in this work were obtained from the VLT/UVES ESO archive for 3 nights in 2001 and 6 nights in 2004 with a spectral resolution of $\sim$42,000 and  have been already published in \citet{2003ApJ...591L..45H,2010ApJ...708L...5M,2012ApJ...751..147H}.   We extracted the raw data from the ESO archive\footnote{Based on observations collected at the European Southern Observatory under ESO programmes 267.C-5736(A) and 074.C-0604(A).} and reduced them with the ESO UVES pipelines under the ESO Recipe Flexible Execution Workbench environment \citep{2013A&A...559A..96F}. We obtain 21 spectra that cover both \OIa\ and H$\alpha$ lines. The 2001 December 14 and 21 and the 2004 December 16--18 (UT) spectra were taken during  Star A eclipse phase, the spectrum acquired on 2001 December 29 was out of eclipse, and the spectra obtained on 2004 December 13--15 were approaching  Star A eclipse phase \citep{2010ApJ...708L...5M,2012ApJ...751..147H}.

We removed  telluric absorption and subtracted photospheric features near the \OIa\  and H$\alpha$ lines using a  K7 PMS template (see e.g., \citealt{2018ApJ...868...28F} for details). We adopted a heliocentric radial velocity $v_{\rm helio}$=18.676\kms\ as the the systemic velocity \citep{2010ApJ...708L...5M,2012ApJ...751..147H}. This value is consistent with the error-weighted mean radial velocity (19.4$\pm$0.1\kms) of other NGC~2264 cluster members \citep{2016ApJ...821....8K}. We shifted the  \OIa\ lines to the systemic velocity and present the spectrum with the highest S/N, the one obtained on 2001 December 21, in Fig.~\ref{Fig:meanline_profile}. Individual H$\alpha$ and \OIa\  profiles are shown in  Figs.~\ref{Fig:line_profile_Halpha} and \ref{Fig:line_profile}. 

 The \OIa\ emission shows the same double-peaked profile {\rev at} all observing epochs.  Note that the sharp blue-shifted spikes in some profiles are due to sky contamination because: (1) they are truly spikes, i.e.  their width is much narrower than the spectral resolution ($\sim$7\kms); (2) sky lines peak near these spikes and, as they get stronger, so {\rev do} the spikes; (3) for four spectra taken in 2001 covering also the \OIb\ line, we detect the same strong and sharp spikes we see in the \OIa\ profiles. The {\rev background} H$\alpha$ line, produced mostly by the H{\scriptsize II} region (Cone Nebula) around KH~15D, is also present but the ratio of {\rev its} intensity to the source emission is much smaller than  the sky lines at the \OIa\ emission. This explains why we do not see such spikes in the reduced H$\alpha$  profiles (see Fig.~\ref{Fig:line_profile_Halpha}).

 The \OIa\ profiles for KH15D differ from those found in most accreting  T~Tauri stars (TTS). In TTS with higher accretion luminosity the \OIa\ profiles typically show both a high velocity component (usually but not exclusively blueshifted) attributed to a jet and a low velocity component (LVC) which is also often blueshifted \citep{1995ApJ...452..736H}. The low velocity component is found in a broader range of TTS, including those with low accretion luminosity and transition disks, and itself can consist of both a broad and a narrow component, which have been attributed to slow disk winds from the inner or outer disk, respectively \citep{1995ApJ...452..736H,2016ApJ...831..169S,2018ApJ...868...28F,2019ApJ...870...76B}. In some cases, the low velocity component is centered on the stellar velocity, and may be bound material on the surface of the disk,  see e.g. Fig.~24 in \citet{2016ApJ...831..169S}. The low velocity components are always single peaked and have line widths that correlate with disk inclination, indicative of broadening by Keplerian rotation.

The \OIa\ profile in KH 15D is distinctive from other TTS. The profile shows symmetric red and blue peaks\footnote{We have mirrored and the subtracted the original spectrum in Fig.~1 and find no asymmetry greater than three times the rms next to the line.} at radial velocities of $\pm$20~\kms\ and looks similar over multiple epochs \footnote{ Note that the claimed factor of $\sim 2$ decrease in the \OIa\ flux  on 2001 December 20 \citep{2010ApJ...708L...5M,2012ApJ...751..147H} is likely spurious. The ESO archive does not contain any spectrum acquired on that date. Instead, most likely, the correct spectrum is the one obtained on December 21st, the line profile is the same as  in  \cite{2010ApJ...708L...5M}. Because KH~15D brightened by 0.77 magnitude in the $R$ band from  December 20 to 21 \citep{2014AJ....147....9W} and this magnitude is used to convert the line equivalent width into flux, an incorrect assignment of the R band can explain the factor of $\sim 2$ underestimation in the \OIa\ flux.}. While such low velocities are characteristic of TTS low velocity components,  no other low velocity components to date have shown a double peaked structure. A jet close to the plane of the sky can have low radial velocities, as can be inferred for some TTS where line ratios for several forbidden lines reveal properties characteristic of shocked, rather than thermally excited, gas for a few sources that would be classified as LVC based on their kinematic properties \citep{2018ApJ...868...28F}.  This is the interpretation proposed by \cite{2010ApJ...708L...5M}, that the \OIa\ profile from KH~15D arises in a pair of identical approaching and receding jets in the plane of the sky. {\rev However, in this interpretation we would expect to see variable profiles at different epochs as noticed in other TTS \citep{2016ApJ...831..169S}. Therefore, we put forward} an alternate possibility, informed by recent work on TTS forbidden lines, that the \OIa\ traces the surface of a disk and is broadened by Keplerian rotation. Its double peaked structure, unlike other LVC in TTS, would result from the fact that the disk is more like a ring, rather than extended beyond $\sim$5~au, as required to fill in the central dip in most TTS \citep{1995ApJ...452..736H,2016ApJ...831..169S}.

\section{H$\alpha$ profiles}\label{decom}

 An extensive analysis of H$\alpha$ profiles and associated emission variability is provided in \cite{2012ApJ...751..147H}. The analyzed spectra covered five contiguous observing seasons (from 2001 to 2006) over which star A was fully visible as well as partially occulted (in ingress and egress events) and fully occulted. Some of these spectra are the same from which we have extracted the \OIa\ profiles and are presented in Fig.~\ref{Fig:line_profile_Halpha}  in order to  emphasize the difference in the line profiles of H$\alpha$ and \OIa.

The out of eclipse and ingress spectra (left panel) are characterized by an inverse P-Cygni-type profile with a strong redshifted absorption whose changes in centroid velocity can be explained by an accretion stream onto A \citep{2012ApJ...751..147H}. 
The eclipse and egress spectra (right panel) present a more symmetric profile with often more pronounced broad extended wings (up to several hundreds
\kms ), which are characteristic {\rev of}  actively accreting stars. Additional spectra in this phase can be seen in  Fig.~9 of \cite{2012ApJ...751..147H}.  \cite{2010ApJ...708L...5M} interpreted the eclipse profiles as tracing the same bipolar jet as \OIa\ since H${\alpha}$ also shows red and blue peaks, although the blue/red intensity is quite variable and the velocities somewhat higher than \OIa. Alternatively, we might still be seeing accretion onto A (and perhaps A+B) through reflected light from the back of the wall of the circumbinary disk. This could explain the similarity of some ingress and mid-eclipse H$\alpha$ profiles and has been proposed by \cite{2012ApJ...751..147H} to explain the presence of the broad wings during these phases.

While a detailed analysis of the H$\alpha$ profiles is beyond the scope of this work, we draw the reader's attention to the difference between the H$\alpha$ and  \OIa\ profiles. While the H$\alpha$ profiles, even during eclipse, are mostly asymmetric and variable in flux, the \OIa\ profiles are persistently {\rev more} symmetric through all the phases (Fig.~\ref{Fig:line_profile}) and show only minor variations in flux (Fig.~1 in \citealt{2010ApJ...708L...5M}).
 This is different from the \OIa\ emission tracing jets which shows strongly variable line intensities and profiles as the emission arises in time variable knots of shocked gas (e.g., \citealt{2011ApJ...736...29H}, \citealt{2016ApJ...831..169S}.) Therefore, our preferred explanation is that the two lines trace different phenomena, with the \OIa{}, being more stable,  probing the surface of the gaseous circumbinary disk.

\section{A circumbinary gaseous ring} \label{sec:discussion}

Here, we explore the possibility that the \OIa\ emission originates from the surface of a circumbinary  gaseous disk and use the line profile to constrain its size. In the assumption of Keplerian broadening, the $\sim$40\,\kms\ separation  of the two peaks points to a characteristic [O {\scriptsize I}] emitting radius of $\sim$3\,au,   consistent with the  mean circumstellar  dust ring radius inferred from modeling the photometric variability of KH~15D \citep{2004ApJ...607..913C}. 

To model in more depth the entire line profile, we adopt a power-law distribution for the O~{\scriptsize I} surface density: $\Sigma({\rm O}$~{\scriptsize I})$\propto r^{\alpha}$, where $r$ is the radial distance from the star. We assume optically thin emission and an  edge-on ($i=90^{\circ}$) disk for simplicity, since (1) photometric variability points to such high disk inclination \citep{2002PASP..114.1167H} and  (2) the \OIa\ line profile is {\rev nearly} symmetric with two peaks (see Fig.~\ref{Fig:meanline_profile}). The line modeling includes thermal broadening for a temperature of 5,000~K  \citep{2016ApJ...831..169S} and convolution  with the instrumental width of $\sim$7\,\kms. We vary $\alpha$ from $-4$ to 4, the inner disk radius ($r_{\rm in}$) from 0.05 to 2.55\,au and outer radius ($r_{\rm out}$) from $r_{\rm in}$ out to 13\,au to perform a grid search and find the parameters leading to the lowest reduced $\chi^2$. We focus the fit on  the \OIa\ line profile obtained during the night of 2001 December 21 because of its high S/N.  We further normalize the observed line profile to its peak and fit it with modelled line profiles that are similarly normalized to their peaks. From this grid search we find the following best-fit parameters:  $\alpha=-2.9$, $r_{\rm in}$=0.57\,au, and $r_{\rm out}$=5.2\,au. We also implement a Markov Chain Monte Carlo (MCMC) procedure \citep{2017ARA&A..55..213S} to find the best-fit parameters and evaluate their uncertainties. The posterior distributions with the best-fit parameters and their uncertainties are shown in Fig.~\ref{Fig:Corner}. These values are consistent with those obtained through the grid search. Note that this best fit profile to the 2001 December 21 spectrum reproduces rather well all the other profiles (see Fig.~\ref{Fig:meanline_profile} and ~\ref{Fig:line_profile}), again highlighting the temporal stability of the \OIa\ emission.

\begin{figure*}
\begin{center}
\includegraphics[width=1.5\columnwidth]{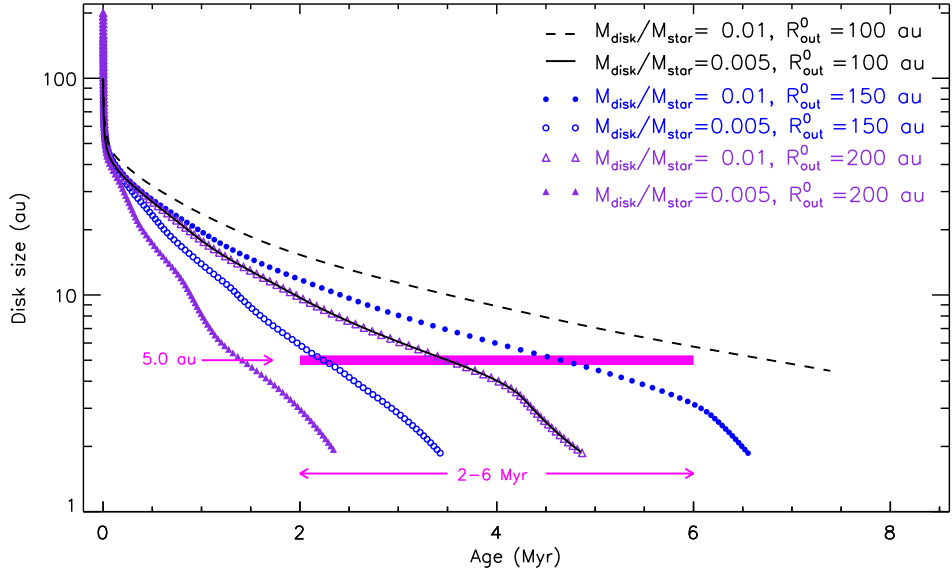}
\caption{The sizes of disks around a 1.5~$M_{\odot}$ young star decrease with the ages due to the external photoevaporation within the UV strength of 4470~$G_{0}$, like KH~15D. The different symbols  are different starting disk masses and sizes.}\label{Fig:size}
\end{center}
\end{figure*}

It is also worth noting that the \OIa\ surface brightness distribution follows a rather steep power law: $I\approx r^{-2.8}$, in comparison to  other disks \citep{1995ApJ...452..736H,2016ApJ...831..169S}. As pointed out in \citet{1995ApJ...452..736H}, a steep power law index of $-$3 is expected if the emission is proportional to the flux from the central star and the disk is flat. Thus, our modeling results further support a disk origin. Interestingly, the inner and outer  gas disk radii inferred from the  [O {\scriptsize I}] emission are very close to those  of the dust ring derived from light curve modeling (see also next Section). These results together point to a disk with a relatively large inner radius and a very small outer radius, a circumbinary ring of gas. 

 \cite{2012A&A...547A..68M} investigated the effect of high-energy stellar photons on the disk structure and the gas chemical composition by varying the X-ray radiation from 0 to 10$^{32}$~erg~s$^{-1}$ and the FUV from 10$^{29}$ to 10$^{32}$~erg~s$^{-1}$. Their calculations show that there exists a radially extended hot ($\sim$5000\,K) region on top of the disk surface\footnote{from $z/r\sim$0.2 at r$\sim$0.6~au to $z/r\sim$0.5 at r$\sim$5~au, where $z$ is the disk vertical height and $r$ is the radial distance from the star} with hydrogen densities of $\sim10^{6}-10^{8}$ cm$^{-3}$. The \OIa\ line should be tracing this region.

 \cite{2010ApJ...708L...5M} claim that there is a slight blue asymmetry (14\%) in the \OIa\ line profiles. For a system like the KH~15D with a close binary, the [O {\scriptsize I}] emitting regions can be disturbed by the time varying potential of the binary system  \citep{2017A&A...604A.102T}, causing some variability and asymmetry in the \OIa\ line profile as proposed by \cite{2010ApJ...708L...5M}. However, as discussed in Sect.~\ref{sec:data}, the blue part of \OIa\ line profile is also contaminated by sky lines, hence changes restricted to a narrow range of velocities close to the sky lines should be interpreted with caution.  Excluding these, likely spurious, features we do not identify any asymmetry in the highest S/N \OIa\ profile.

\section{Inner and outer truncation of the disk}

Several theoretical papers have shown that  the inner region of a disk surrounding a close binary is quickly dissipated (e.g., \citealt{1993prpl.conf..749L,1994ApJ...421..651A}). {\rev Recently, \citet{2008MNRAS.391..815P} carried out numerous simulations, using the test particle approah, to derive a simple relation  between the gap radius and binary semi-major axis ($a$), eccentricity ($e$), and primary/companion mass ratios ($q$):} $R_{\rm gap}\approx 1.93a(1+1.01e^{0.32})[q(1-q)]^{0.043}$. For the KH~15D system, $a \sim$\,0.25~au, $e \sim$\,0.6, and $q \sim$\,0.83--0.97 \citep{2006ApJ...644..510W,2018AJ....155...47A}, thus the disk inner radius is expected to be at $\sim$0.8\,au {\rev according to these simulations}. {\rev However, as shown in the more detailed physical models of  \cite{2016ApJ...827...43M} and \cite{2019ApJ...871...84M} , it is actually difficult to define the cavity radius for an eccentric binary, as gas from the circumbinary disk flows in generating complex and transit streams. Therefore, we consider our} estimate of $\sim 0.5$\,au from modeling the \OIa\ line {\rev pretty close to the possible cavity radius}.

\cite{2004ApJ...607..913C} argued that the disk of KH~15D cannot extend beyond $\sim$5\,au to maintain rigid precession and explain the long-term photometric variability of the system (but see \citealt{2013MNRAS.433.2157L} for the possibility of the disk extending to 5--10\,au). Recently, \cite{2016AJ....151...90A}  confirmed the rigid precession of the ring and estimated a velocity (across the sky) of $\sim$15\,m~s$^{-1}$ for the projected edge of the ring. Then, the precession rate of the ascending node of the ring can be   estimated to be 0\fdg18~yr$^{-1}$/$\bar{a}$, where $\bar{a}$ is mean radius of the circumbinary ring in $au$. Using the approximate expression relating the precession rate of  ascending node of an inclined circumbinary ring to the mean ring radius from \cite{2004ApJ...607..913C}, the masses (0.715 and 0.74\,\Msun\ \citealt{2018AJ....155...47A}) and orbit (0.25\,au, \citealt{2004ApJ...603L..45W}) of the binary, the mean radius of the ring should be $\sim$4~au. This value is also {\rev close to the characteristic \OIa\ emitting radius, see Sect.~4.}

Preventing the outer boundary from viscous spreading may require some sort of outside confinement. 
\cite{2004ApJ...607..913C} proposed a planet exterior to the ring. Here, we explore  external photoevaporation as an alternative.  At a projected separation of $\sim$37$''$, south-west  from KH~15D, there is a massive star HD~47887 ($T_{\rm eff}$=24000$\pm$1000K, \citealt{2014A&A...562A.143F}). Based on \textit{Gaia} Data Release 2, KH~15D and HD~47887 have  similar parallaxes (1.2691$\pm$0.0751\,mas vs. 1.3272$\pm$0.0751\,mas). Thus, it is very likely that both of them belong to the  NGC~2264 cluster. With a distance from us of $\sim$750\,pc, the projected distance between them is $\sim$0.135~pc.

Taking {\rev this projected distance,} the model atmosphere from \cite{2007ApJS..169...83L}, and the broad-band photometry from \cite{2001KFNT...17..409K}, the UV field strength near KH~15D produced by HD~47887 is $\sim$4470~$G_{0}$\footnote{G0 is the Habing unit of UV radiation corresponding to the integrated flux (1.6$\times10^{-3}$~erg$^{-1}$~cm$^{-2}$~s$^{-1}$) over the wavelength range 912\AA\ to 2400\AA\ used in \cite{2018MNRAS.481..452H}.}. Under such UV field strength (4470~$G_{0}$), we make a toy model to assess the timescale for dissipating the outer disk of KH~15D as close in as 5\,au. In the models, the initial disk sizes are set to be 100, 150, or 200~au, disk masses are 0.0075 or 0.015~$M_{\odot}$ (0.5\% or 1\% of the KH~15D system mass), and the inner disk radius is fixed to be 0.1~au. The disk material has  a surface density $\Sigma\propto r^{-1}$, and the central binary is simplified  as one single star with a mass of 1.5\,\Msun. Furthermore, we assume that the bulk of the mass loss due to external photoevaporation is driven from the disk outer edge, as  in \cite{2018MNRAS.481..452H}, and neglect the disk viscous evolution. We interpolate the disk mass-loss rates within the UV field strength of 4470~$G_{0}$ during the disk dissipation using the FRIED grid of mass-loss rates for externally irradiated protoplanetary disks \citep[see details in][]{2018MNRAS.481..452H}. The mass-loss rates range from $\sim$10$^{-6}$ to $\sim$10$^{-10}$\Msunyr\ when disk size decrease from hundreds\,au to 5\,au.
The resulting disk sizes from our toy models are shown in Fig.~\ref{Fig:size} as a function of age.  Our calculations show that the size of a disk within the UV field strength of 4470~$G_{0}$ can decrease to 5~au at the age of the cluster \citep[2--6\,Myr,][]{2009MNRAS.399..432N,2016ApJ...831..116L}. Thus, photoevaporation can explain the outer truncation of the disk.

\section{conclusion}

We provide a new interpretation to the double-peaked \OIa\ emission line as arising from the {\rev hot} surface of the gaseous disk surrounding the spectroscopic binary KH~15D. Line profile modeling  constrains the emission to a relatively narrow {\rev radial extent} $\sim$0.5--5\,au, {\rev similar to the extent of the dust ring} inferred from modeling the source photometric variability. The relatively large inner edge  is likely set by the binary dynamical interaction. We show that the small outer disk radius could be shaped by external photoevaporation driven by  UV photons from the nearby massive star HD~47887.  {\rev Newer high signal-to-noise \OIa\ line profiles would be useful to further test our interpretation and detect any subtle changes that might arise from disk precession.}

\acknowledgments
We thank the anonymous referee for comments that helped to improve this paper and K. Kratter for inputs on the theory of circumbinary disks.  I.P. and S.E. acknowledge support from a Collaborative NSF Astronomy \& Astrophysics Research Grant (ID: 1715022 and ID:1714229). This material is based upon work supported by the National Aeronautics and Space Administration under Agreement No. NNX15AD94G for the program "Earths in Other Solar Systems". The results reported herein benefitted from collaborations and/or information exchange within NASA Nexus for Exoplanet System Science (NExSS) research coordination network sponsored by NASA's Science Mission Directorate.

%



\appendix
\section{Comparison of observed \OIa\ line profiles and the model line profile}\label{app}
\begin{figure*}
\begin{center}
\includegraphics[width=1\columnwidth]{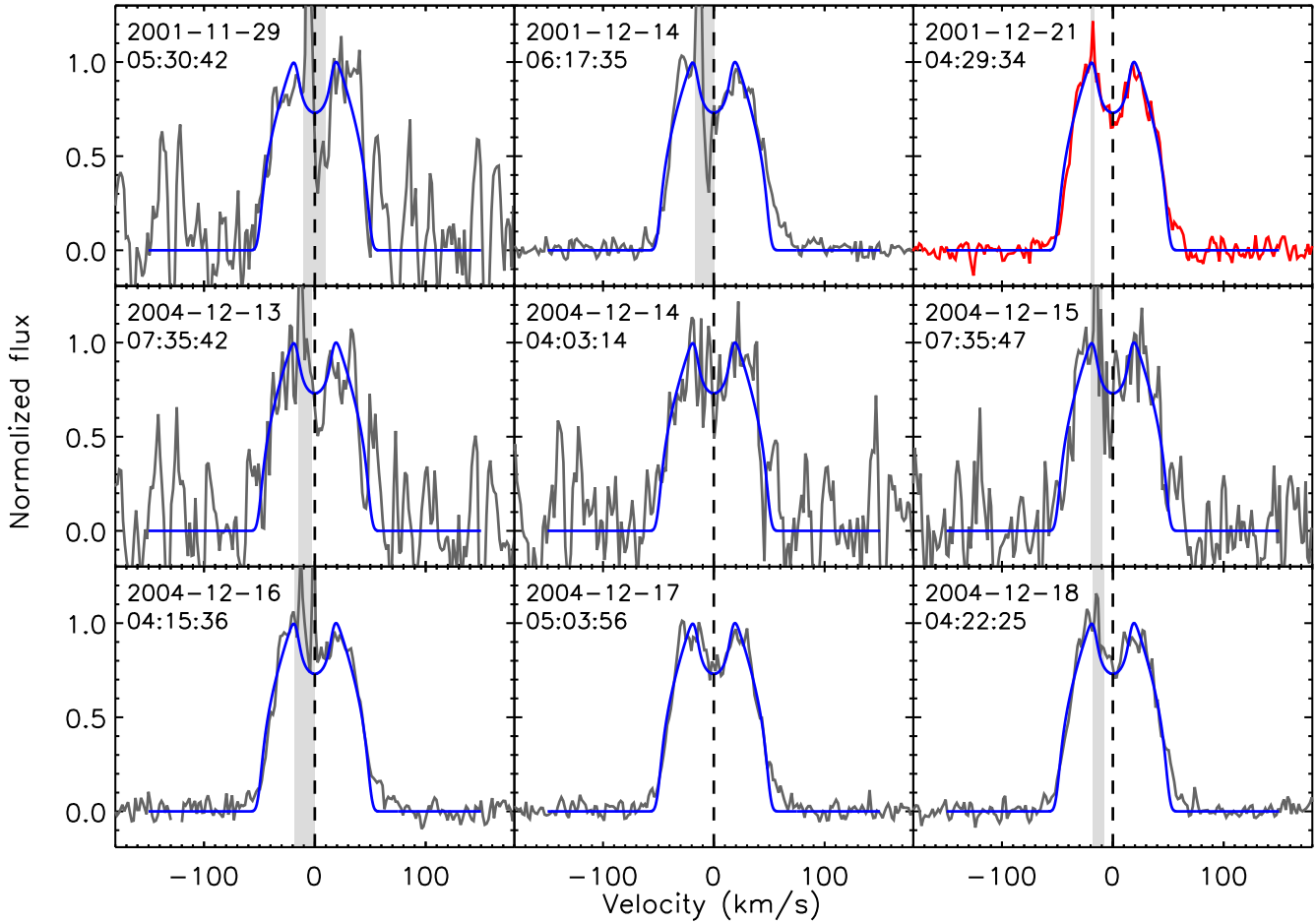}
\caption{\OIa\ line profiles (grey and red lines) obtained between 2001 and 2004, compared to our best-fit line profile (blue lines), see Appendix~\ref{app} for details. The red line  shows the \OIa\ line profile taken on 2001 December 21 and used for the fitting. For multiple observations during 2001 December 14 and 2004 December 13-16, we only show one spectrum per night since there are no detectable variations in \OIa\ profiles.  The zero velocity marked with the dash line is that of the KH~15D  rest frame. The grey-color filled region marks wavelength ranges where sky contamination is likely.  }\label{Fig:line_profile}
\end{center}
\end{figure*}

Figure~\ref{Fig:line_profile} shows the comparison of the line profiles observed at different epochs (grey lines) with the best-fit model line profile (blue lines).  The latter is obtained by fitting the 2001 December 21 profile (outermost right panel). We chose this line profile because (1) it has the highest S/N, and (2) it shows the least contamination from  sky lines. The best fit parameters are: $\alpha=-2.82$, $r_{\rm in}$=0.54\,au, and $r_{\rm out}$=5.0\,au, see  also \sect~\ref{sec:discussion}.

\end{document}